# COSMIC BACKGROUND ANISOTROPIES IN CDM COSMOLOGY

NAOSHI SUGIYAMA

*Departments of Astronomy and Physics, and Center for
Particle Astrophysics, University of California, Berkeley, CA 94720*

*Department of Physics, Faculty of Science,
University of Tokyo, Tokyo 113, Japan*

## ABSTRACT

Cosmic microwave background (CMB) anisotropies and density fluctuations are calculated for flat cold dark matter (CDM) models with a wide range of parameters, i.e., $\Omega_0, h$ and $\Omega_B$ for both standard recombination and various epochs of reionization. Tables of the power spectrum of CMB anisotropies in the form of $C_\ell$'s as a function of $\ell$ are presented. Although the Harrison-Zeldovich initial spectrum is assumed in these tables, we present simple approximations for obtaining the $C_\ell$'s corresponding to a tilted spectrum from those with a Harrison-Zeldovich spectrum. The $\sigma_8$ values are obtained for the matter density spectrum, with $\sigma(10°)$, fixed $Q_{rms-PS}$ and COBE DMR 2 year normalizations. Simple modifications of the fitting formula of the density transfer function which are applicable for models with high baryon density are given. By using both numerical results and these fitting formulae, we calculate the relation between $\sigma_8$ and $Q_{rms-PS}$, and find good agreement. Velocity fields are also calculated.

*Subject headings*: cosmology: cosmic microwave background - dark matter

e-mail: sugiyama@pac2.berkeley.edu

# I. Introduction

An increasing number of recent cosmic microwave background (CMB) anisotropies experiments on various angular scales (see e.g., White, Scott and Silk 1994) is providing new and important information about the creation and formation of the universe. The combination of CMB anisotropy experiments with large scale structure data enables us to test the consistency of structure formation models. Only a few groups have developed Boltzman codes to calculate detailed density and temperature perturbations in an expanding universe, and produced CMB anisotropy spectra for different models. Now that the observational precision is increasing, the variance in the results is dominated by uncertainties in galactic foreground on the observational side, and by subtle differences in the computations and model parameters, on the theoretical side. Although the treatment of perturbations by each group is quite similar, the detailed results differ slightly. It is not even sure that all calculations produce precisely the same results for the same models. Stompor (1994) has compared his results with several other works. He found agreement with most of groups to within about 20%. Holtzman (1989) has obtained temperature power spectra and matter power spectra for various cosmological models. However his temperature spectra do not contain the entire information necessary for observational comparisons on all scales. The main purpose of the present paper is to present a catalog of the temperature and density spectra for different cosmological models appropriate to the post-COBE era.

During the past decade, what is practically a standard structure formation model has emerged as the (biased) cold dark matter (CDM) model with $\Omega_0 = 1$, $\Omega_B h^2 \simeq 0.01$ and $h = 0.5$, where $h$ is the non-dimensional Hubble constant normalized to 100km/s/Mpc (see e.g., Ostriker 1993). However the accumulation of data from large-scale structure surveys reveals the weak point of the standard CDM model. Since more power on large scales is needed, just from the shape of the density power spectrum, we should set the CDM shape parameter $\Gamma \equiv$ $\Omega_0 h \simeq 0.25$ (Efstathiou, Bond, and White 1992; Peacock and Dodds 1994).

Moreover, the normalization of CMB anisotropies by COBE DMR detector causes another problem for this model. There is excessive power for the $\Omega_0 = 1$ standard model at $8h^{-1}$Mpc. Assuming a COBE normalization, the standard model has to be not merely unbiased but rather anti-biased. Again, a low density or a low value of the hubble constant seems to be preferable. Do we have to take either a low value of $h$ or a low value of $\Omega_0$? Recent surveys of Cepheids in two Virgo cluster galaxies (Pierce et al. 1994; Freedman et al. 1994) show evidence for a high value, i.e., $h \simeq 0.8$. Together with the cosmic age problem, this may require low density universe models. However if we are living in a big hole which behaves like an open universe, the local value of Hubble constant does not have to be the global value (Bartlett et al. 1994). If we once know the precise value of the Hubble constant, we probably also know the baryon density $\Omega_B$ from Big Bang primordial Nucleosynthesis; $0.01 < \Omega_B h^2 < 0.015$ (Walker et al. 1991). While this prediction is fairly robust, recent direct measurement of deuterium abundance in Ly-$\alpha$ clouds (Songaila et al. 1994), if confirmed, may indicate a lower value of $\Omega_B$. Moreover, the baryon abundance in the Coma cluster (White et al. 1993) requires either low $\Omega_0$ or high $\Omega_B$. It may be premature to give any conclusive arguments about cosmological parameters.

In this *paper*, we consider CDM models with a wide range of parameters, i.e., $h, \Omega_B$ and $\Omega_0$. Assuming adiabatic perturbations with a Harrison-Zeldovich power spectrum as initial conditions, we provide CMB anisotropies and matter power spectra of these CDM models. It is possible to obtain CMB spectra with tilted initial spectra from models with Harrison-Zeldovich spectra by using the simple approximation presented in §3. Throughout this *paper*, we only consider a universe with flat geometry. In other words, we assume the existence of the cosmological constant for low density models. Open universe models are also one of the interesting possibilities as a candidate for large-scale structure formation (see e.g., Coles and Ellis 1994). Although there are several recent attempts to obtain a power spectrum from the inflation scenario in open universes (Lyth and Stwart 1990; Ratra and Peebles 1994; Bucher, Goldhaber and Turok 1994), however, the precise shape of the density power spectrum beyond the curvature scale in an open universe has not been yet definitively determined. Since the CMB anisotropies on COBE scales are strongly dependent on the behaviour of the power spectrum around the curvature scale, we do not include open universe models in this paper. For specific shapes of power spectra, CMB anisotropies and large scale structure constraints are shown in Kamionkowski, Spergel and Sugiyama (1994), Kamionkowski et al. (1994), and Gorski et al. (1994b).

## II. Calculations and Assumptions

We employ the gauge invariant method (Bardeen 1980; Kodama and Sasaki 1984) for our numerical calculations of perturbations. We note that the final results for the density and CMB spectra could not depend on the gauge choice. The reason for choosing the gauge invariant formalism is mainly convenience, i.e., easiness of setting an initial condition and absence of unphysical gauge dependent modes. Here we only calculate scalar perturbations. In other word, we do not take into account the tensor (gravitational wave) mode which may dominate on very large scales. Because this tensor mode is added by quadrature sum in temperature power spectrum, independent calculations are possible. We also neglect polarization which is only important for reionized universe models or on very small scales for standard recombination.

We will show the outline of our calculations. Detailed treatments of perturbations are shown in Sugiyama and Gouda (1992). We set adiabatic initial conditions at $T = 10^7$K when the radiation components are totally dominated against matter components. Before the electron mean free time becomes comparable to the expansion time which usually occurs at the standard recombination era, we treat the photons and baryons as a single viscous fluid. Then we start to solve a Boltzmann equation for the photons by expanding in multipole components. Three massless species of neutrinos are followed by another Boltzmann equation from the beginning. We calculate these coupled equations up to the present epoch at which the photon temperature $T_0$ is 2.7K.

Treatment of the recombination process is one of the most important issues to consider in calculating CMB anisotropies, since the detailed behavior of photon diffusion damping (Silk 1968) is strongly dependent on the time evolution of the free electron number density. Here we solve the ionization history by following Peebles (1968) and Jones and Wyse (1985). We take the mass fraction of helium to be $Y = 0.23$. The recombination process of helium is not considered because of simplicity. Throughout our calculations, helium is treated as a neutral atomic gas. The recombination process of helium takes place well before recombination process of hydrogen. At that time, photon and baryon are tightly coupled. Therefore ignoring helium recombination does not cause any significant difference.

In this paper, we consider not only standard recombination but also reionization of hydrogen atoms after recombination. This reionization does not affect the evolution of the matter density perturbations for CDM models because the recombination epoch is well within the matter-dominated regime. On the other hand, CMB fluctuations are suppressed on small scales and produced on the horizon scale of the new last scattering surface (Sugiyama, Silk and Vittorio 1993).

As our end results, we obtain matter fluctuations $\Delta(\eta_0, k)$ and multipole components of temperature fluctuations $\Theta_\ell(\eta_0, k)$ for each $k$ mode. Here $\eta_0$ is the conformal time at the present epoch. For the matter spectrum, we obtain the transfer function $T(k) \equiv \left(D(\eta_i)/D(\eta_0)\right) \left(\Delta(\eta_0, k)/\Delta(\eta_i, k)\right)$, where $D(\eta)$ is the growth factor of density fluctuations at $\eta$ and $\eta_i$ is some very early time. Although the full information about CMB anisotropies is contained in a similar transfer function in $k - l$ space (Hu and Sugiyama 1995b), we integrate over all $k$ modes in order to obtain the observed quantities (Bond and Efstathiou 1987) in the form

$$\frac{2\ell + 1}{4\pi} C_\ell = \frac{V}{2\pi^2} \int \frac{dk}{k} \frac{k^3 |\Theta_\ell(\eta_0, k)|^2}{2\ell + 1}, \qquad (2.1)$$

where $V$ is the volume of the fundamental cube. Then the rms temperature fluctuations are expressed as $(\Delta T/T)^2_{rms} = \sum_\ell (2\ell+1) C_\ell W_\ell / 4\pi$ with $W_\ell$ as the experimental window function. In order to obtain $C_\ell$, we have to set the initial power spectrum shape. We take the Harrison-Zeldovich initial condition in the results given below.

Since this $\Theta_\ell$ is a highly oscillatory function in $k$ for large $\ell$, we need a huge number of steps in $k$ in order to obtain a smooth power spectrum. For example, if we require $C_\ell$ up to $\ell = 1000$, we must take at least 3000 or more steps in $\log k$. Since both computational time and disc space are limited, however, 3000 steps are not realistic. Instead of taking this large a number of $k$ steps, we smooth the final $C_\ell$ obtained by 500 steps for models with standard recombination and 100 steps for ones with reionization using the Wiener filter method. We find almost perfect agreement with a 3000 step computation.

## III. Results

### 3.1 Models

Here we take about 50 different combinations of cosmological parameters. For each combination, we consider both standard recombination and reionization after recombination as alternative thermal histories. The reionization epoch $t_*$ is determined by requiring the optical depth

$$\tau_* = \int_{t_*}^{t_0} dt x_e n_e \sigma_T \tag{3.1}$$

to be unity. Here $t_0$ is the present time, $x_e$ is the ionization fraction of electrons, $n_e$ is the total electron number density and $\sigma_T$ is the Thomson scattering cross section. As for standard $\Omega_0 = 1$ CDM models, we take $h = 0.3, 0.5,$ and $0.8$, and $\Omega_B = 0.01, 0.03, 0.06, 1.0$ and the Big Bang nucleosynthesis (BBN) value $0.125 h^{-2}$. For cosmological constant-dominated models ($\Lambda$ CDM), we set $h =$ 0.5 and 0.8, and $\Omega_B = 0.01, 0.03, 0.06$ and the BBN value. We consider $\Omega_0 = 0.1, 0.2, 0.3$, and 0.4. The resultant $\ell(\ell+1)C_\ell/2\pi$ are presented in Tables 1 to 11. We also show the most plausible models from large-scale structure formation which have $\Gamma \equiv \Omega_0 h = 0.25$ with the BBN value of $\Omega_B$ in Table 12. We take $\Omega_0 = 0.3, 0.4$, and 0.5 for this case. Moreover we add $\Omega_0 = 0.2$ and $h = 1.0$ which deviate slightly from this region of parameter space. It should be noticed that the $\Omega_0 = 1, h = 0.3$ model shown in Table 1 is also in the preferred region of parameter space. In these tables, we normalize the CMB anisotropies to the COBE 10 degree rms fluctuations of $30\mu K$ by assuming a gaussian window function $W_\ell = \exp\left(-(\ell+1/2)^2\sigma^2\right)$, where $\sigma = 10/2\sqrt{2\ln 2}$ degrees.

### 3.2 Normalization

The normalization of CMB anisotropies is still ambiguous in the present situation. From the COBE first year data (Smoot et al. 1992),

$$Q_{rms-PS}/T_0 = \sqrt{\frac{5}{4\pi}C_2} \qquad (3.2)$$

is required to be $15\mu K$ for the $n = 1$ single power-law spectrum. On the other hand, Gorski et al. (1994) analyzed the two year data by Bayesian power spectrum estimates and obtained $19.9\mu K$ and $20.4\mu K$ with and without the quadrupole anisotropy, respectively, for this power spectrum. By using the same 2 year data, Bennett et al. (1994) claimed $Q_{rms-PS} = 17.6\mu K$ from a likelihood analysis of the cross correlation function and Wright et al. (1994) obtained $19.3\mu K$ in a maximum likelihood analysis of their Monte Carlo simulations. A further complication is that the 2-year released data are processed in ecliptic coordinates, and give a normalization approximately $1\mu K$ higher than the data set in galactic coordinates that has been analyzed by the COBE DMR team (Bunn, Scott and White 1995). Although the differences are not significant beyond the $1\sigma$ level, they are enough to provide ambiguity in the final results. We may have to wait until the 3rd and 4th year data of COBE DMR are released to obtain a definitive normalization. Moreover, as for the fixed $Q_{rms-PS}$ normalization, we can only employ this method for the models whose $\ell(\ell+1)C_\ell$'s are close enough to flat in $\ell$ over large scales, as is expected for the $n = 1$ single power-law power spectrum. Flat $\Omega_0 = 1$ models have a nearly flat tail. However, $C_\ell$'s for $\Lambda$ CDM models show structure on small $\ell$'s (see e.g., Sugiyama and Silk 1994). This structure prevents us from simple use of the fixed $Q_{rms-PS}$ normalization. Strictly speaking, we should analyze the full DMR data set to obtain the correct normalization. Bunn and Sugiyama (1995) applied Bayesian analysis on the DMR 2 year data. They obtained the maximum likelihood value of $Q_{rms-PS}$ for $\Lambda$ CDM models. We can read off the normalization from their results. We include their maximum likelihood number of $Q_{rms-PS}$ for $h = 0.5$ and $\Omega_B = 0.03$ in Figure 1. This maximum likelihood number is used as the DMR 2 year normalization in this *paper* with ignoring small dependence on $h$ and $\Omega_B$. This may not be still the final number, however, because the likelihood functions do not have sharp peaks for low density models. Moreover these models should be renormalized if the gravitational wave mode is included.

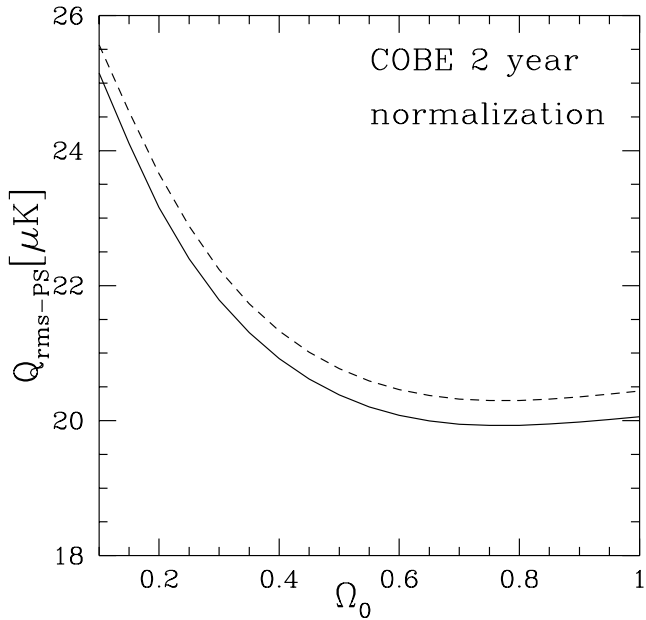

Figure 1: Values of the most likely $Q_{rms-PS}$ of flat CDM models with the $n = 1$ initial spectrum, $h = 0.5$ and $\Omega_B = 0.03$ as a function of $\Omega_0$. Bayesian analysis of COBE 2 year data (Bunn and Sugiyama 1995) is employed. Solid and dashed lines are the cases including and excluding quadrupole data, respectively.

Therefore in these tables of $C_\ell$'s, we are rather taking the simplest normalization method than possibly more realistic ones. We normalized the CMB aniso-tropies to the DMR 10 degree fluctuations, i.e., so-called $\sigma(10°)$ normalization as mentioned in the previous section. This corresponds to $Q_{rms-PS} = 15\mu K$ for the $n = 1$ single power-law. The $Q_{rms-PS}$'s for this normalization are shown as a function of $\Omega_0$ in Figure 2. Readers are invited to renormalize each set of models for themselves. By using Figure 1 and reading $C_2$'s from the tables, it is easy to obtain DMR 2 year–normalized $C_\ell$'s.

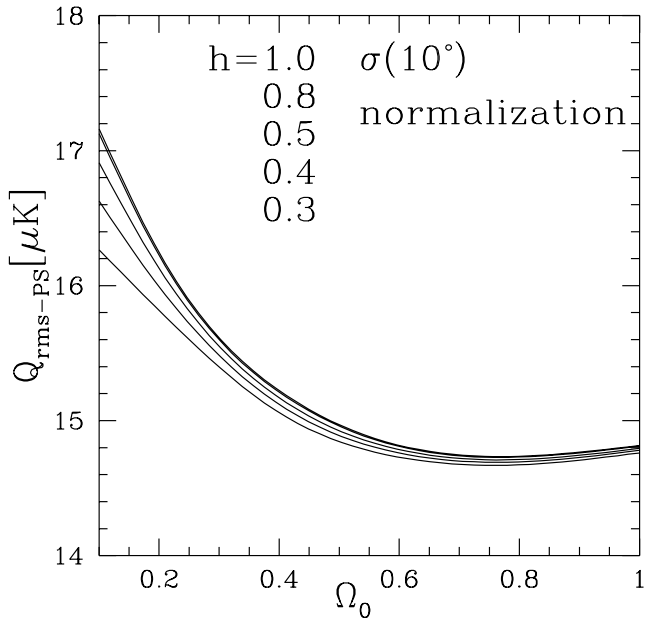

Figure 2: $Q_{rms-PS}$'s of $n = 1$ flat CDM models normalized to $\sigma(10°)$ as $30\mu K$. From the top to the bottom, $h = 1.0, 0.8, 0.5, 0.4$ and $0.3$. We take $\Omega_B = 0.0125h^{-2}$ which is BBN value. Changing $\Omega_B$ doesn't change these lines very much.

## 3.3 Tilted Spectrum

Deviation from scale invariance is associated with the contribution of the gravitational wave. Davis et al. (1992) present the relation between the spectral index and the ratio of scaler and tensor contribution to the quadrupole: $T/S = 7(1-n)$. While corrections of this relation may be required (Liddle and Lyth 1992; Kolb and Vadas 1994), considering tilted spectra is still essential to investigate the tensor mode.

Although we assume a Harrison-Zeldovich spectrum in these tables, it is rather simple to get approximate $C_\ell$'s with a tilted spectrum near $n = 1$ from

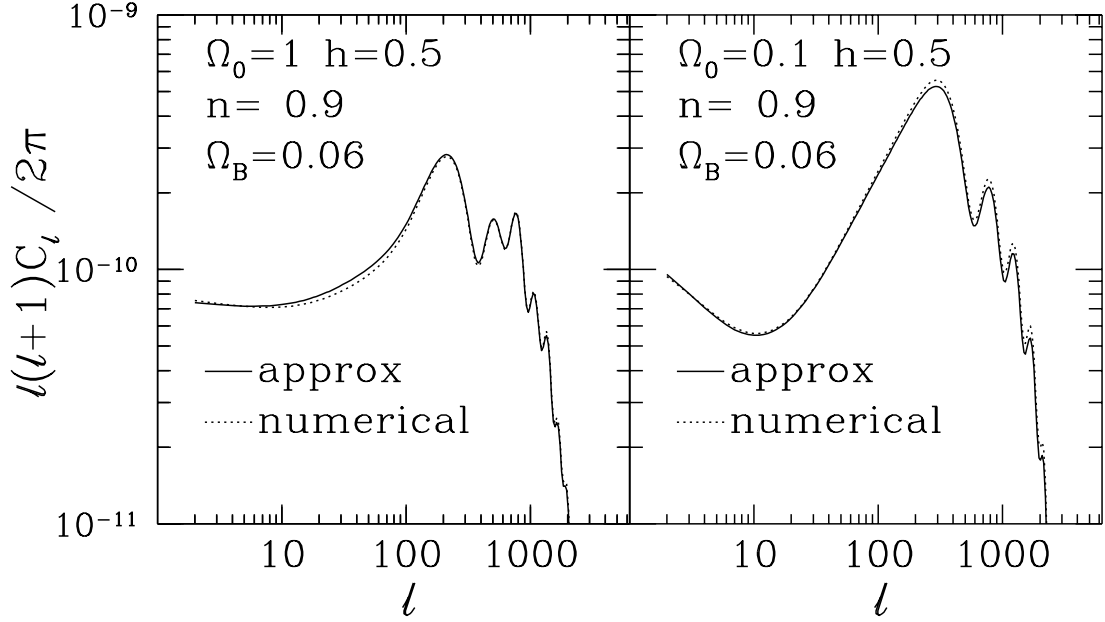

Figure 3: $C_\ell$'s with $n = 0.9$ tilted spectrum normalized to $\sigma(10°)$ as $30\mu K$. The left panel is $\Omega_0 = 1$ and right is $\Omega_0 = 0.1$. Dotted lines are numerical results and solid lines are analytically obtained from $C_\ell$'s with $n = 1$ by employing equation (3.4).

these tables. We can express $C_\ell$'s of arbitrary spectral index $n$ as

$$\frac{2\ell+1}{4\pi} C_\ell = \frac{V}{2\pi^2} \int \frac{dk}{k} \frac{k^3 |\bar{\Theta}_\ell|^2}{2\ell+1} \left(\frac{k}{k_*}\right)^{n-1}, \qquad (3.3)$$

where $\bar{\Theta}_\ell$ are the temperature anisotropies with the $n = 1$ spectrum and $k_*$ is an arbitrary normalization of wave number. Expanding $(k/k_*)^{n-1}$ around $n = 1$ as $1 + (n-1)\ln(k/k_*)$ and assuming that most of the contribution to $\Theta_\ell(\eta_0, k)$ is coming from $k = \ell/\eta_0$, we obtain the relation, apart from normalization, as

$$C_\ell = (1 + (n-1)\ln(\ell/\ell_*)) \bar{C}_\ell, \qquad (3.4)$$

where $\ell_* \equiv k_* \eta_0$ and $\bar{C}_\ell$ refers to the $n = 1$ spectrum. Since this expansion is only appropriate around $\ell_* = \ell$, we should take this to be 100. In Figure 3, we show $C_\ell$'s for $n = 0.9$ obtained by this approximation for both $\Omega_0 = 1$ and $\Lambda$ CDM models, together with real numerical results. On all scales, errors are less than about 10% in power even for the $\Lambda$ CDM model.

### 3.4 Reionization

From the Gunn-Peterson test, we know the universe should reionize at some epoch after recombination at least by $z \simeq 5$ (Gunn and Peterson 1965; Steidel and Sargent 1986; Jenkins and Ostriker 1991). Physical mechanisms of reionization are considered by various authors (see e.g., Tegmark, Silk and Blanchard 1994; Fukugita and Kawasaki 1994). Although CDM models seem to be incapable of reionizing before $z \sim 100$ from these works, we can consider more realistic reionization at relatively late epochs. If the reionization epoch $z_*$ is much larger than $1/\Omega_0$, we obtain a useful approximation for the relation between $z_*$ and optical depth $\tau_*$ of equation (3.1) as

$$z_* = 100 \tau_*^{1/3} \Omega_0^{1/3} \left( \frac{0.025}{x_e \Omega_B h} \right)^{2/3}. \tag{3.5}$$

Hence $\tau_* = 1$, as taken in our tables, is about the maximum possible reionization. Even if we consider later reionization, i.e., $\tau \lesssim 1$, the CMB spectrum can be significantly modified (Sugiyama, Silk and Vittorio 1993).

Here we show the $\tau$ dependence of $C_\ell$. In Figure 4(a), we plot numerical results for the $C_\ell$'s with $\tau = 0$ (no reionization), $0.5, 0.8, 1.0, 1.5$ and $2.0$ for $\Omega_0 = 1$, $h = 0.5$ and $\Omega_B = 0.05$. These $C_\ell$'s are also presented in Table 13. It is shown that reionization smooths out the original temperature fluctuations on scales smaller than a few degrees ($\ell \geq 70$). Even for $\tau_* = 0.5$, damping on small and intermediate scales is significant. It is known this damping is proportional to $\exp(-\tau)$ in temperature from diffusion damping (Hu and Sugiyama 1994). In our numerical calculations, we obtain the correct amount of damping on small scales. On the other hand, new fluctuations on larger scales that correspond to the horizon scale at last scattering are created by Doppler motions of electrons against photons in the case of larger $\tau_*$.

Now, let us propose a simple interpolation and extrapolation of $C_\ell$'s with any $\tau_*$'s from two sets of data, i.e., $\tau_* = 0$ and $1$, which are presented in our tables

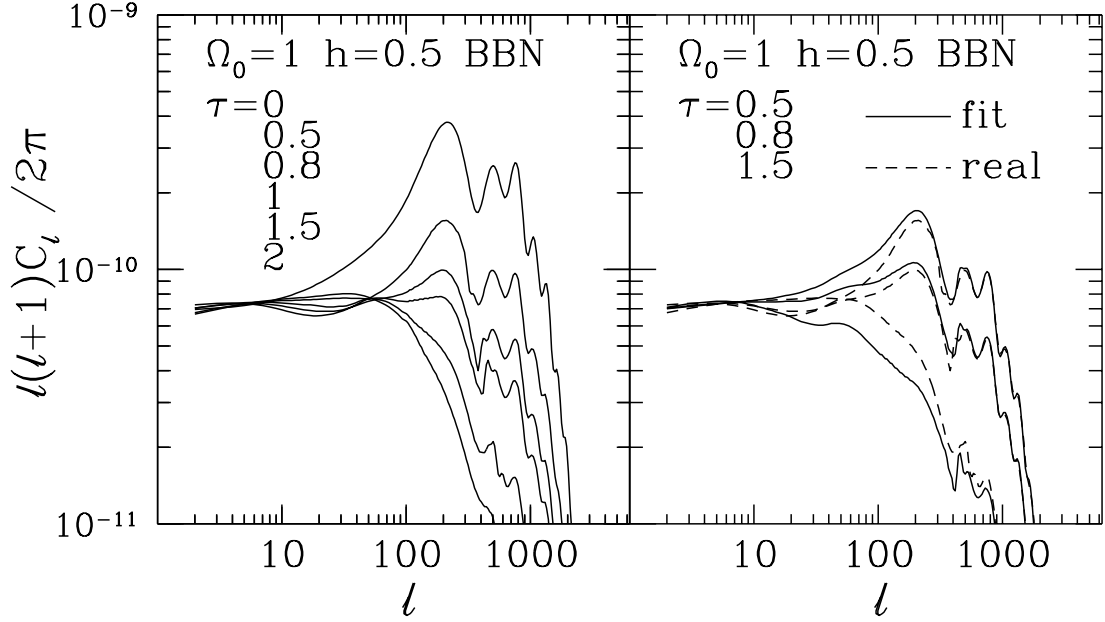

Figure 4: $C_\ell$'s of $n = 1$ reionized models with the no reionization model. We take $\Omega_0 = 1$, $h = 0.5$ and $\Omega_B = 0.05$ which is BBN value. We use $\sigma(10°)$ normalization. In the left panel, the optical depth $\tau$ is taken as 0 (no reionization), $0.5, 0.8, 1, 1.5$ and 2 from the top to the bottom at $\ell > 100$. Dashed lines of the right panel are same numerical $C_\ell$'s of the left panel with $\tau = 0.5, 0.8$ and $1.5$. Solid lines are interpolated or extrapolated from $\tau = 0$ and 1 data by using the equation (3.6).

as:

$$C_\ell(\tau) = (C_\ell(1)/C_\ell(0))^\tau C_\ell(\tau). \tag{3.6}$$

We show this fit compared with numerical calculations in Figure 4(b) for $\tau = 0.5, 0.8$ and $1.5$. The fit works very well on small scales but is not perfect on the scales corresponding to the new last scattering horizon. However the maximum difference is about 30% in power, i.e., 15% in temperature which is not unacceptable for determining the tendency of the dependence on ionization and for providing rough constraints on reionization models from observations even on a few degree scales.

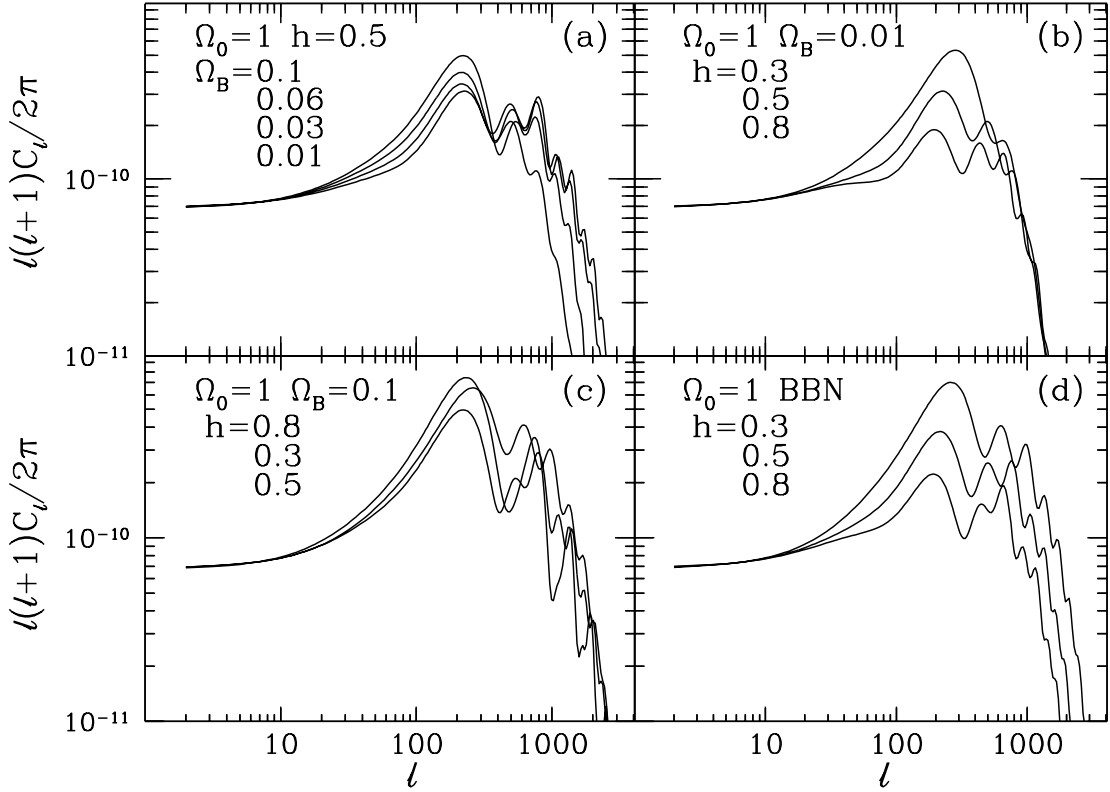

Figure 5: $C_\ell$'s of $\Omega_0 = 1$ and $n = 1$ models with different cosmological parameters; (a) changing $\Omega_B$, (b) changing $h$ with low $\Omega_B$, (c) same as (b) with high $\Omega_B$ and (d) same as (b) with $\Omega_B h^2$ fixed to be 0.0125. The order of captions on each panels correspond to the order of lines at the first peaks. $\sigma(10°)$ normalization is taken.

### 3.5 Parameter Dependence

In Figure 5, the $\Omega_B$ and $h$ dependences of the $\Omega_0 = 1$ models are shown. All of the following physical interpretations have been presented in Hu and Sugiyama (1995a, b). Decreasing $\Omega_B h^2$ increases the sound speed of the baryon-photon fluid and prevents the fluctuations from growing. Hence smaller $\Omega_B h^2$ means lower peaks. This tendency is shown in Figure 5a. There is another factor entering in the $h$ dependence. Increasing $h$ causes the epoch of matter-radiation equality to be earlier. This makes the gravitational potential deeper on scales larger than the sound horizon. Generally speaking, adiabatic growth is bigger for deeper potential because of gravitational infall. In case of small $\Omega_B$, however, the sound speed is still high even after matter-radiation equality and the window when adiabatic growth dominates is small. Moreover the deeper potential also means that there is a larger Sachs-Wolfe effect (Sachs and Wolfe 1967) which is the red-shift. This red-shift by the Sachs-Wolfe effect cancels out the blue-shift by gravitational infall. Therefore the relative height of the peaks becomes lower with increasing $h$ as shown in Figure 5b. It is clearer if we fix $\Omega_B h^2$, i.e., the sound speed, as shown in Figure 5d. In fact, the deeper potential with fixed sound speed means the lower peak. On the other hand, if $\Omega_B$ is fixed at an intermediate value, these two effects, i.e., decreasing the sound speed and increasing the depth of the potentials, are competitive. This is why Figure 5c shows apparently strange behaviour.

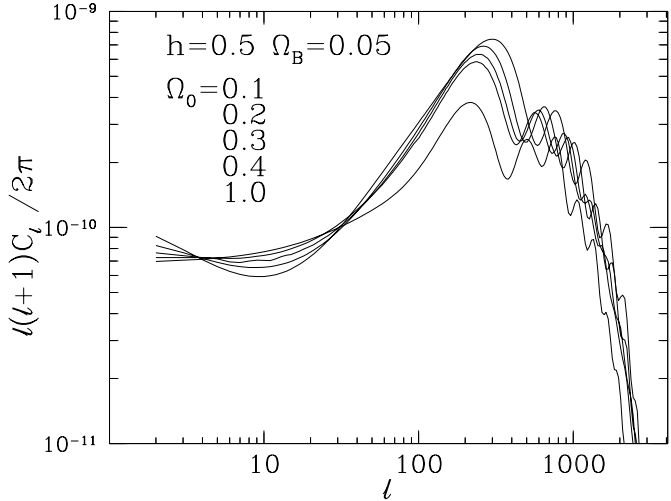

Figure 6: $C_\ell$'s of models with $n = 1$ and different $\Omega_0$'s normalized to $\sigma(10°)$. We fix $h$ and $\Omega_B$ as 0.5 and 0.05 respectively. From the top to the bottom at first peaks, $\Omega_0 = 0.1, 0.2, 0.3, 0.4$ and $1.0$.

From Figure 5, we can also infer the $\Omega_B$ and $h$ dependences of the diffusion damping scale. Although the diffusion damping scale in $\ell$ is proportional to $\Omega_B^{1/2} h^{1/2}$ from the simple assumption of neglecting the recombination process, the numerical results do not quite follow this fitting because the true damping scale is strongly dependent on the recombination process. In these figures, the $h$ dependence is very weak while the $\Omega_B$ dependence is about that expected from the analytic estimates.

We show the $\Omega_0$ dependence for fixed $\Omega_B h^2$ in Figure 6. There are two significant results. One is the existence of the minimum around $\ell = 10$ for low density models and the other is the $\Omega$ dependence of the peak height. Both of these features can be explained by the Sachs-Wolfe effect caused by decaying of gravitational potentials (Sachs and Wolfe 1967). Here we refer to this effect as the integrated Sachs-Wolfe (ISW) effect to distinguish it from the usual Sachs-Wolfe (SW) effect which is caused by the difference of gravitational potential between the last scattering surface and the present epoch. Only after the cosmological constant becomes comparable to the matter energy density, is this ISW effect important on large scales. Hence the 'thickness' of this effect is quite large. Because of the thickness cancellation, this effect is damped with increasing $\ell$. Therefore minima appear for low density models. On the other hand, around the location of the peak, the ISW effect is important right after recombination. If we take a small value of $\Omega_0 h^2$, the universe is not completely matter-dominated even at the recombination epoch. The gravitational potential is still decaying on intermediate scales that are larger than the maximum sound horizon and smaller than the scales which cross the horizon at full matter domination. Hence we expect an excess in the temperature spectrum on scales larger than the first peak as shown in Figure 6. Even the first peaks shown in this figure are not the simple adiabatic peaks whose locations are expected to be on much smaller scales from the projection of the maximum sound horizon. Because of adding this ISW effect, the first peaks of low density models become higher and move to larger scales. Eventually, the first peak locations are nearly independent of $\Omega_0$ for flat CDM models. On the other hand, for low density open models, locations of peaks including the first peaks significantly move to smaller scales because of the deviations of geodesics (Kamionkowski, Spergel and Sugiyama 1994). It may be possible to use the second or third peak locations as the indicators of $\Omega_0$ for $\Lambda$ CDM models.

Figure 7: Contours of $\sigma_8$ on $\Omega_0 - h$ plane. We employ $\sigma(10°)$ normalization. Bold solid lines are $\sigma_8 = 1$. As one moves up the right hand side, $\sigma_8$ increases 0.1 at each contour. For some panels, we also plotted the contour with $\sigma_8 = 0.05$. Panels (a), (b), (c) and (d) are models with $\Omega_B = 0.01, 0.03, 0.06$ and $0.0125 h^{-2}$, respectively.

### 3.6 Matter Power Spectrum

We now discuss the matter power spectrum. One of the current key observed quantities is the amplitude of mass fluctuations at $8h^{-1}$Mpc, i.e, $\sigma_8$ which is defined as

$$\sigma_R^2 = <|\delta M/M(R)|^2> \equiv \frac{V}{2\pi^2} \int \frac{dk}{k} k^3 |\Delta(\eta_0, k)|^2 \left(\frac{3j_1(kR)}{kR}\right)^2 . \qquad (3.7)$$

In Figure 7, contour plots of $\sigma_8$ in the $\Omega_0 - h$ plane are shown for $\Omega_B = 0.01, 0.03$, 0.06 and the BBN value with $\sigma(10°)$ normalization. Although it is possible to convert these values into fixed $Q_{rms-PS}$-normalized values or the DMR 2 year normalization by using Figures 2 or 1, we also show the same contour plots with $Q_{rms-PS} = 20\mu K$ normalization and the DMR 2 year (including quadrupole data) normalization in Figures 8 and 9, respectively.

Figure 8: Same of figure 7 but $Q_{rms-PS}$ is fixed to be $20\mu K$.

If the baryon density can be neglected, we find that our matter transfer functions themselves are well described by a fitting formula from Bardeen et al. (1986):

$$T(k) = \frac{\ln(1 + 2.34q)}{2.34q}[1 + 3.89q + (16.1q)^2 + (5.46q)^3 + (6.71q)^4]^{-1/4}, \quad (3.8)$$

where $q = k/[\Omega_0 h^2 \text{Mpc}^{-1}]$. Following Holtzman's calculations (Holtzman 1989), Peacock and Dodds (1994) modify this $q$ in order to take into account effects of baryon density by writing $q = k/[\Omega_0 h^2 \exp(-2\Omega_B)\text{Mpc}^{-1}]$. We may have to point out that $\Gamma$ which is measured in their paper should be defined as $\Omega_0 h \exp(-2\Omega_B)$ instead of $\Omega_0 h$. We verified that this works remarkably well for $\Omega_0 = 1$ models. However, further modification is needed for low density universe models because the effect of baryons should be larger for these models. The natural choice of the parameter is $\Omega_B/\Omega_0$. Here we introduce the scaling

$$q = k/[\Omega_0 h^2 \exp(-\Omega_B - \Omega_B/\Omega_0)\mathrm{Mpc}^{-1}]. \qquad (3.9)$$

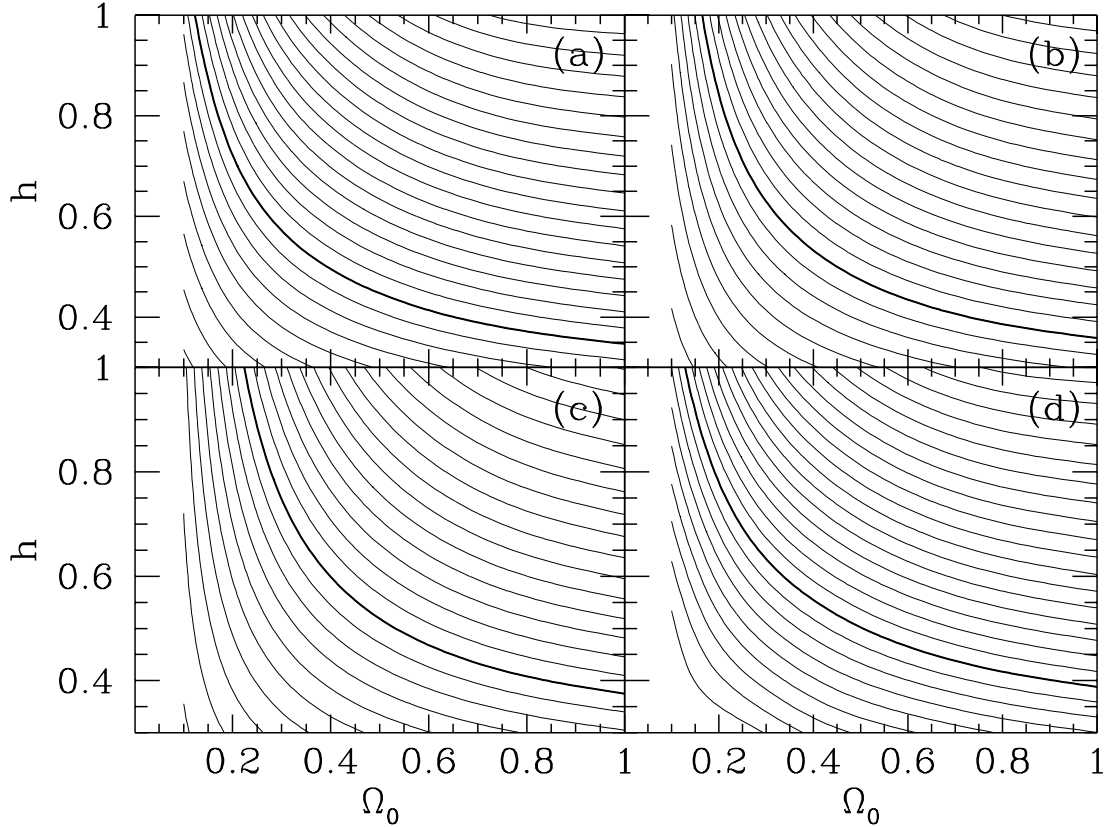

Figure 9: Same of figure 7 but we employ COBE DMR 2 year normalization (see Figure 1).

In Figure 10, we show that this simple modification works quite well. We can only follow the cases where $\Omega_B$ is no larger than 50% of $\Omega_0$. Even in this extreme case, however, our fitting gives the right tail on small scales and the right number for $\sigma_8$ with less than 10% error, if we take the same normalization values of the matter power spectrum on very large scales, relative to the numerical results. We could do more further comparisons between numerical results and this analytic fitting formula. For the quadrupole anisotropy, we can assume most of contributions are coming from the Sachs-Wolfe effect. If the universe is matter-dominated, there is no ISW effect in $\Omega_0 = 1$ models on large scales. After the cosmological constant dominates relative to matter density for $\Lambda$ CDM models, the ISW effect becomes important. The solution of the Boltzmann equation is written for the quadrupole moment as

$$\Theta_2(\eta_0, k)/5 = \frac{1}{3}\Psi(\eta_d, k) j_2\left(k(\eta_0 - \eta_d)\right) + 5 \int_{\eta_d}^{\eta_0} \frac{d\Psi(\eta, k)}{d\eta} j_2\left(k(\eta_0 - \eta)\right) d\eta, \quad (3.10)$$

where $\Psi$ is the gravitational potential, $j_2$ is the second order spherical Bessel function, and $\eta_d$ is the conformal time at the last scattering surface. The first term on the right hand side is the SW effect and the second one is the ISW effect. From the Poisson equation, we get the relation between $\Psi$ and $\Delta$ as

$$\Psi(\eta,k) = -\frac{4\pi G\rho a^2}{k^2}\Delta(\eta,k)$$
$$= -\frac{3}{2}\frac{(Ha)^2}{k^2}\frac{D(\eta)}{D(\eta_0)}\Delta(\eta_0,k)\,, \qquad (3.11)$$

where $\rho$ is the total energy density, $a$ is the scale factor, and $H$ is the Hubble constant. The growing mode solution $D$ is described in the linear regime (see, e.g., Peebles 1980) as

$$D(\eta) = H\int_0^\eta \frac{d\eta}{H^2 a}\,. \qquad (3.12)$$

Figure 10: Transfer function of total density perturbations for $\Omega_0 = 0.1$ and $\Omega_B = 0.0125h^{-2}$. Left and right panels are models with $h = 0.5$ and $0.8$, respectively. Solid lines are numerical results. Dashed and dotted lines are the fitting formula of Bardeen *et. al* (1985) parametrized by equation (3.9) and by the Peacock and Dodds, respectively.

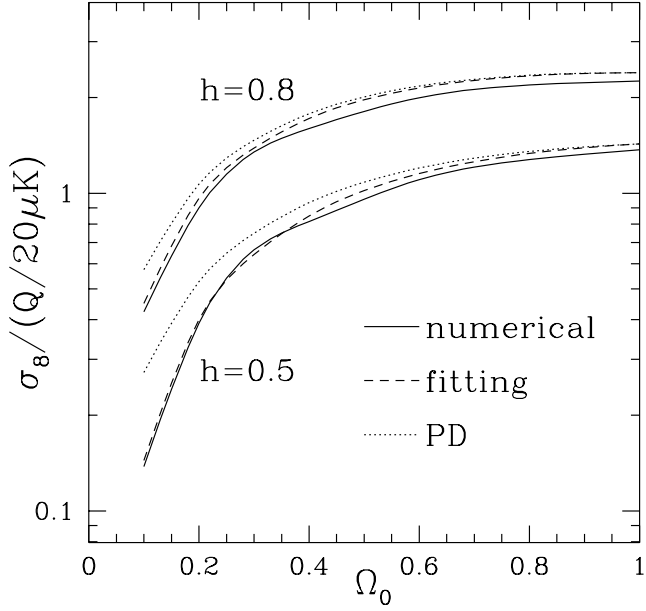

Figure 11: The ratio of $\sigma_8$ and $Q_{rmp-PS}/20\mu K$ for $h = 0.5$ and $h = 0.8$ with $n = 1$ and $\Omega_B = 0.0125 h^{-2}$. Solid lines are numerical. Dashed and dotted lines are obtained from the fitting formula parameterized by equation (3.9) and by the Peacock and Dodds, respectively.

From these equations, we can analytically calculate the ratio of $Q_{rms-PS}$ and $\sigma_8$ by employing the fitting formula for matter fluctuations. In Figure 11, the number of $\sigma_8$ normalized to $Q_{rms-PS}$ to be fixed as 20 $\mu K$ is shown for $h = 0.5$ and 0.8, BBN models as a function of $\Omega_0$. We get very good agreement with numerical values for analytic values with our fitting formula.

We may have to replace the condition $\Gamma = 0.25$ with $\Omega_0 h \exp(-\Omega_B - \Omega_B/\Omega_0)$ = 0.25. However the difference between our modified transfer function and that of Peacock and Dodds is only important in the case of $\Omega_B \gtrsim 0.1\Omega_0$ and low density. If we assume the BBN value of $\Omega_B$, this happens only for small $h$.

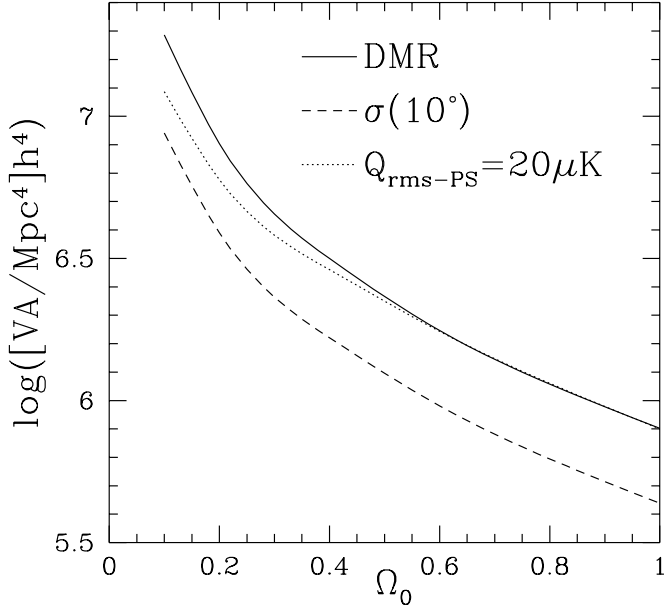

Figure 12: The normalization of present matter power spectrum $A$ times volume factor $V$ and $h^4$ as a function of $\Omega_0$. Solid, dashed and dotted lines are obtained from DMR 2 year, $\sigma(10°)$ and fixed $Q_{rms-PS}$ ($20\mu K$) normalizations, respectively.

Moreover, if we also take the large scale structure value $\Gamma = 0.25$, deviations of these two transfer functions only appear if $\Omega_0 \gtrsim 0.5$ and $h \lesssim 0.5$. Therefore, only the very small window of $\Omega_0$ for the BBN value with $\Gamma = 0.25$ cannot be probed using Peacock and Dodds' transfer function.

Finally, we plot the normalization of present matter power spectrum $A$ times $V$ and $h^4$ as a function of $\Omega_0$ for $\sigma(10°)$, fixed $Q_{rms-PS}$ and DMR 2 year (including quadrupole data) normalizations in Figure 12. This normalization $A$ is defined as

$$|\Delta(\eta_0, k)|^2 \equiv A k T(k)^2 . \qquad (3.13)$$

### 3.7 Velocity Field

Next we will show another observational quantity, i.e., the velocity field. In numerical calculations, we could directly obtain the velocity field. The expected velocity field at distance $r$ is

$$v(r)^2 \equiv \frac{V}{2\pi^2} \int \frac{dk}{k} k^3 |v(\eta_0, k)|^2 W(kr)^2 , \qquad (3.14)$$

where $v(\eta_0, k)$ is the present velocity perturbation in Fourier space, and $W(kr)$ is

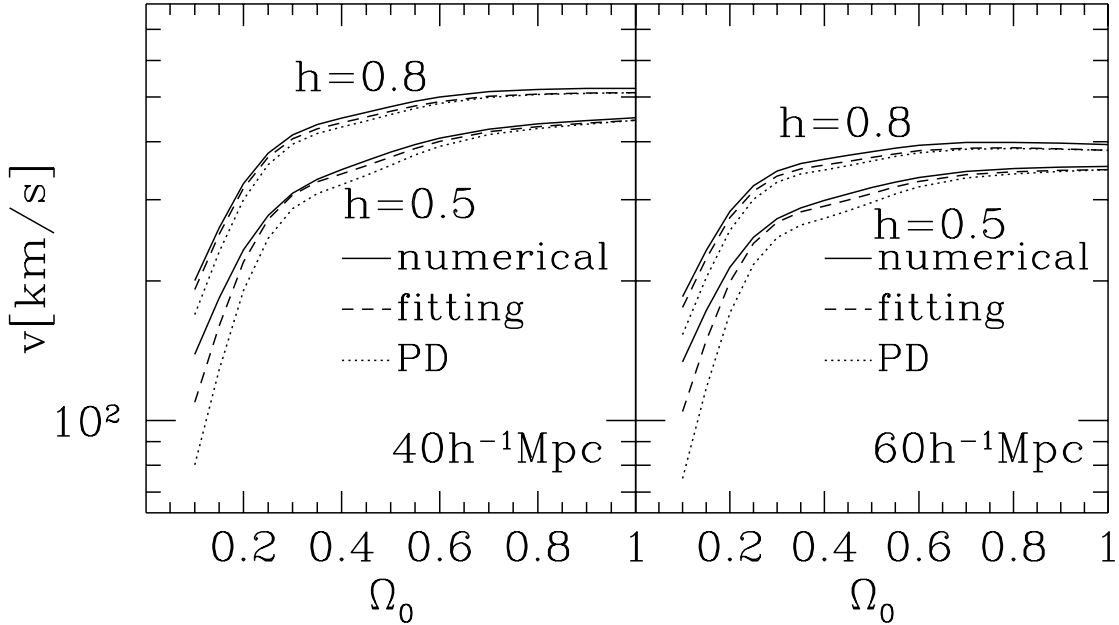

Figure 13: The velocity fields which are smoothed by a $12h^{-1}$Mpc gaussian window function and 40 (the left panel) or $60h^{-1}$Mpc (the right panel) of the top hat window function. The model parameters are $n = 1$ and $\Omega_B = 0.0125h^{-2}$. We take $h = 0.5$ and 0.8. DMR 2 year normalization is employed. Solid lines are numerical. Dashed and dotted lines are obtained from the fitting formula parameterized by equation (3.9) and by Peacock and Dodds, respectively.

the Window function. We take here the top hat window function smoothed over $r_s = 12h^{-1}$Mpc by a Gaussian window function following the Potent analysis (Bertschinger et al. 1990) as $W(kr) = \left(3j_1(kr)/kr\right) \times \exp\left(-k^2 r_s^2/2\right)$. In Figure 13, the expected velocities at $r = 40$ and $60h^{-1}$Mpc normalized to DMR 2 year data (including quadrupole data) are shown as a function of $\Omega_0$. We also plot velocities obtained by fitting a formula to the transfer function by assuming a simple linear relation between density and velocity perturbations (Peebles 1980) as $v = -(Ha/k)f\Delta$ where $f = \ln D/\ln a$. This simple relation is well matched by our numerical calculations. However there are discrepancies between numerical velocities and analytic ones at small $\Omega_0$ for $h = 0.5$. These are because most of the contribution of the velocity fields is coming from around the turning point of density power spectrum where our fitting formula does not work well for high $\Omega_B/\Omega_0$ (see Figure 10). If we plotted $k^3|v|^2$, it varies as $k^2$ on large scales and $k^{-2}\ln k$ on small scales for the $n = 1$ initial spectrum. The turning point corresponds to the

horizon scale at the epoch of matter radiation equality $k_{eq} = 0.064\Omega_0 h^2 \text{Mpc}^{-1}$. Numerically, this turning point is about $k_{turn} = 0.25\Omega_0 h^2 \text{Mpc}^{-1}$. Hence when $r$ is smaller than $k_{turn}^{-1}$, the biggest contribution is always coming from this turning point. However, our transfer function still works better than the previous one by Peacock and Dodds. We also checked the fitting formula of the velocity fields at 40 and $60h^{-1}$Mpc produced by Bond (1994). We found his fitting recovers the results of Peacock and Dodds transfer function but does not give the right answer for the high $\Omega_B/\Omega_0$ case.

Direct comparison between models and observations is made possible by use of the quantity $\beta$ defined as $\beta \equiv f/b_{IRAS}$, where $b_{IRAS}$ is the biasing of IRAS galaxies (see e.g., Strauss and Willick 1994). Simple analytic fitting for $f = \ln D/\ln a$ which is $\Omega^{0.6}$ (Peebles 1980) or modified one for the universe with the cosmological constant which is $\Omega^{0.6} + (1/70)(1-\Omega_0)(1+0.5\Omega_0)$ (Lahav et al. 1991) are provided. Although deviations from numerical values are less than 10% for the former fitting and 5% for the latter one even in case of low $\Omega_0$, we take numerical values here. Observationally, IRAS biasing parameter and optical one are related as $b_{opt} = 1.3 b_{IRAS}$ (Peacock and Doods 1994). Hence we can describe;

$$\beta = 1.3 f/b_{opt} = 1.3 f \sigma_8 \ . \tag{3.15}$$

In Figure 14, we show contours of $\beta$ which are normalized to DMR 2 year data (including quadrupole data).

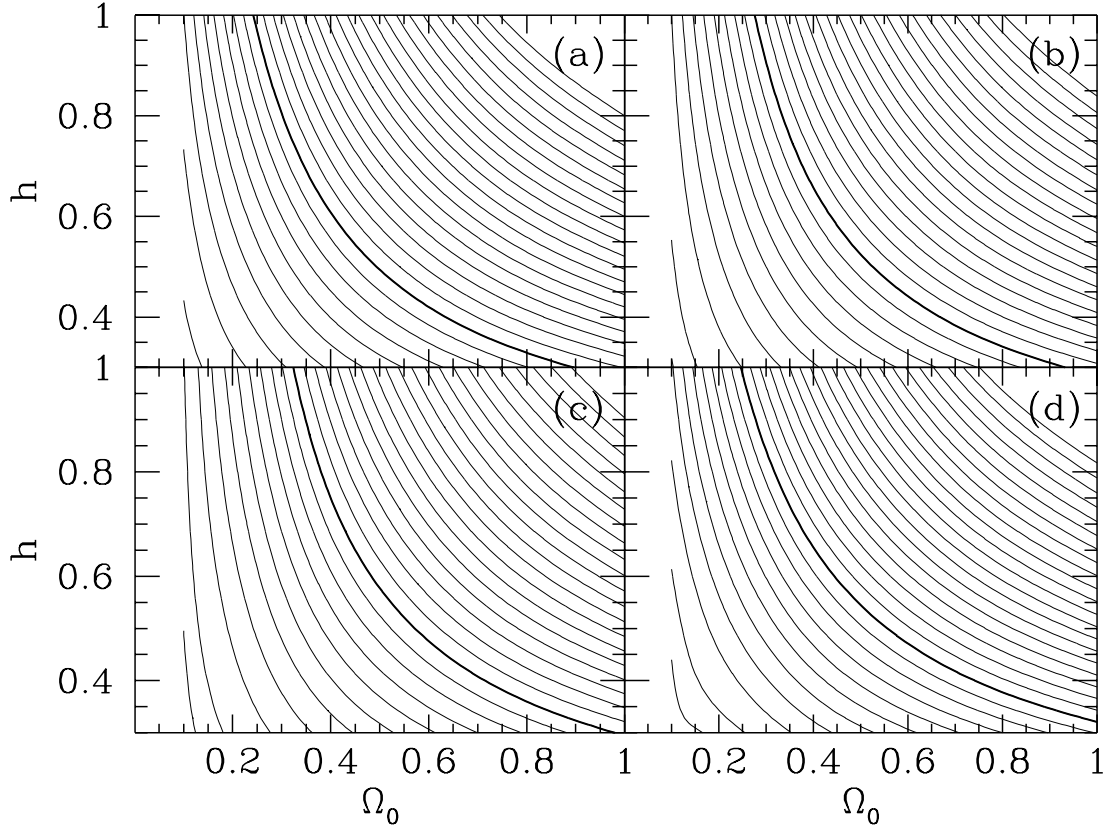

Figure 14: Contours of $\beta$ on $\Omega_0-h$ plane. We employ DMR 2 year normalization. Bold solid lines are $\beta = 1$. As one moves up the right hand side, $\beta$ increases by 0.1 at each contour. For some panels, we also plotted the contour with $\beta = 0.05$. Panels (a), (b), (c) and (d) are models with $\Omega_B = 0.01, 0.03, 0.06$ and $0.0125h^{-2}$, respectively.

## IV. Discussion

We have investigated CMB anisotropies and density fluctuations in CDM models with a wide range of parameters. However we have not attempted to give constraints on specific models in this paper. The current rapid increase in the surveys of large scale structure and CMB anisotropy observations promise to change our insights into structure formation models. Therefore, we have taken not only the most plausible models from the present observations but a large grid of CDM models.

Recently, several alternatives to standard CDM models have been considered.

One is the hot and cold mixed dark matter (MDM) model (Davis, Summers, and Schlegel 1992; Klypin, Holtzman, Primack, and Regos 1993). Introducing an admixture of hot dark matter with the CDM reduces the small-scale power, thereby effectively lowering $\Gamma$ but retaining an $\Omega_0 = 1$ universe. Because models with high massive nutrino mass density are excluded by the observations of damped Ly-$\alpha$ systems (Kauffmann and Charlot 1994), the mass density ratio between hot and cold is taken to be 0.2 : 0.8 (Ma and Bertschinger 1994). Although the density power spectrum is modified on small scales, we do not expect big differences in the CMB anisotropy spectrum. The difference between hot and cold appears on scales smaller than the maximum Jeans scale of the massive neutrinos. Since this is smaller than the maximum Jeans scale of photon-baryon fluid, the first peaks which are defined by this maximum Jeans scale are not modified even if we consider the pure hot dark matter model as the most extreme case of MDM models. There might exist small difference on small scales. However the damping of the CMB power spectrum, which is defined by the diffusion process of photon-baryon fluid, wipes out the difference on very small scales. The only difference we expect in the CMB spectrum between CDM and MDM is the height of higher $\ell$ peaks.

Other alternatives are topological defect models, i.e., cosmic string, domain wall or texture. Usually no initial fluctuations are considered in these models. All density and CMB fluctuations which are observed today were generated by the defects. Hence we expect quite different radiation spectra from the CDM models. As for the density spectrum, however, a very similar spectrum to CDM models is obtained. Although no one has yet properly calculated the CMB spectrum beside the Sachs-Wolfe effect (see e.g., Pen, Spergel and Turok 1994), CMB anisotropies on intermediate scales will provide us with the most important information to distinguish defect models from CDM models that we have presented in this paper.

Baryon dominated models with initial isocurvature perturbations (BDM or PIB) are also possible candidates of structure formation models. The shape of CMB anisotropy spectrum is very different from CDM ones although it is strongly dependent on thermal history of the universe (Hu and Sugiyama 1994). Combining with DMR 2 year data, we will see the consistency of models from CMB anisotropies on intermediate scales and density power spectrum (Hu, Bunn and Sugiyama 1994).

If we can obtain whole sky maps of CMB anisotropies as half degree or better angular resolutions by a new satellite in near future, very detailed structures in CMB power spectrum will become important. In this case, the treatment of temperature perturbations in this *paper* may not be accurate enough. Ignoring the polarization may maximally provide about 10% error in power on small scales (private communications by J.R. Bond, P. Steinhardt and M. White). A more precise treatment of recombination process including the difference between electron and photon temperatures, helium recombination process, stimulated process (Sasaki and Takahara 1993) and so on, may be required. We have to pursue the true temperature spectrum.

The author would like to thank E. Bunn, M. Davis, W. Hu, D. Scott, and M. White for valuable discussions. He especially thanks J. Silk for many enlightening discussions and carefully reading the manuscript. He acknowledges financial support from a JSPS Postdoctoral Fellowship for Research Abroad.

# Tables

Tables are available at an anonymous ftp site;
ftp://pac2.berkeley.edu/pub/sugiyama/tables.ps
ftp://pac2.berkeley.edu/pub/sugiyama/tables.tex

A complete version of this paper (uuencoded) is also available at
ftp://pac2.berkeley.edu/pub/sugiyama/sugiyama.uu